\documentclass[a4paper,10.5pt]{article}

\usepackage[margin=1.16in]{geometry}
\usepackage{authblk} 

\usepackage{setspace}
\usepackage{newtxtext} 
\usepackage{newtxmath} 
\usepackage[
    backend=biber,
    style=numeric-comp,
    sorting=none,
    giveninits=true,
    doi=true,        
    isbn=false,
    url=true,
    eprint=false,
]{biblatex}
\AtEveryBibitem{
    \clearfield{urlyear}
    \clearfield{urlmonth}
    \clearfield{urlday}
}

\usepackage{amsmath}
\usepackage{amsfonts}
\usepackage{enumitem}

\usepackage{hyperref} 
\hypersetup{
    colorlinks=true, 
    citecolor=red,  
    linkcolor=orange, 
    urlcolor=blue    
}

\usepackage{graphicx}
\usepackage{multirow}

\usepackage{float}
\usepackage{listings, color}
\usepackage{here}
\usepackage{bm}
\usepackage{booktabs}
\usepackage{physics}
\usepackage{mathtools}
\usepackage{siunitx}
\usepackage{algorithm}
\usepackage{algpseudocode}

\usepackage{placeins}

\title{\huge BVH-Accelerated Ray Tracing for High-Frequency Electromagnetic Backscattering}

\author[1]{Marco Pasquale}
\author[1]{Andong Hu}
\author[1]{Luca Pennati}
\author[1]{Ivy Peng}
\author[1]{Stefano Markidis}

\affil[1]{\fontsize{10.5pt}{11.5pt}\selectfont Department of Computational Science and Technology, KTH Royal Institute of Technology

\vspace{2mm}
\texttt{\{marcopas,andonghu,pennati,bopeng,markidis\}@kth.se}}

\date{}

\begin{document}

\maketitle

\vspace{-8mm}


\begin{center}
\begin{minipage}{0.85\linewidth}
\begin{abstract}
\noindent As computational complexity in electromagnetics increases with frequency, full-wave solvers become computationally infeasible for electrically large problems. To address this limitation, we present a shooting and bouncing rays (SBR) method for efficiently modeling electromagnetic backscattering of metallic objects in the high-frequency regime. The method couples multi-reflection geometrical-optics ray transport with a physical optics surface integral discretized over ray tubes. To reduce the massive ray-surface intersection search space, we use a bounding volume hierarchy (BVH) and organize the computation as a trace–integrate pipeline. The ray tracing generates hit data, and the physical optics integral is evaluated over valid intersections only. Numerical accuracy is controlled through an incident-ray sampling rule that mitigates phase aliasing in the discretized physical optics integration. The method is accelerated on NVIDIA and AMD GPUs and parallelized with MPI. We validate against analytical Mie solutions for a perfectly electrically conducting (PEC) sphere and demonstrate applicability to a complex aircraft geometry for monostatic radar cross-section prediction.

\vspace{1em}
\end{abstract}
\end{minipage}
\end{center}

\vspace{2em}

\hrule
\vspace{2mm}

\noindent\textbf{\textit{Note.}} This manuscript is an extended version of the paper accepted at the 26th International Conference on Computational Science (ICCS 2026).

\vspace{2mm}
\hrule

\vspace{2mm}
\noindent\textbf{\textit{Software.}} The source code is available at: {\small \url{https://github.com/marco-pas/SagittaSBR}}

\vspace{2mm}
\hrule

\vspace{5mm}

\section{Introduction}
\noindent Computational electromagnetics (CEM) includes both numerical and asymptotic methods for predicting electromagnetic fields governed by Maxwell's equations in realistic geometries and materials. It is widely used across applications such as antenna and microwave component design, wireless propagation and channel modeling, radar and remote sensing, and scattering analysis for complex platforms~\cite{taygur_computation_2018,taygur_bidirectional_2018,yun_ray_2015,schiller_gpu_2015,kasdorf_parallel_2024,gomez_accelerated_2023,purcell_combination_2002}. Within this broad spectrum of applications, different modeling regimes call for different tools. Full-wave methods are preferred when resonance, diffraction, or strong coupling dominate, while asymptotic or hybrid approaches can be more efficient when the geometry is large compared to the wavelength and field interactions become strongly directional.

When the characteristic size of a structure $D$ is much larger than the wavelength $\lambda$ (i.e., $D/\lambda \gg 1$), full-wave discretizations must resolve many wavelengths across the geometry, leading to very fine meshes, large linear systems, and often impractical memory and runtime requirements. This challenge is becoming increasingly important in emerging 6G scenarios extending toward sub-terahertz frequencies~\cite{li_exploring_2025}. In the high-frequency limit $\lambda \ll D$, however, many scattering effects are dominated by specular interactions and slowly varying surface currents, motivating asymptotic approaches that replace a global field solve with ray-like transport and coherent accumulation of scattering contributions.

Ray tracing provides a computational framework for this regime. In computer graphics, ray tracing models transport by launching rays, computing intersections with scene geometry, and applying local interaction rules at each hit. In HF electromagnetics, a closely related structure arises. Rays approximate geometrical-optics (GO) propagation and reflection, while electromagnetic boundary conditions determine the reflected field and phase accumulation along a path. The scattered field is then obtained by coherently assembling contributions over the illuminated surface, commonly through a physical-optics (PO) surface-current integral~\cite{mittra_computational_2014}.

Shooting and bouncing rays (SBR) combines these ideas by \emph{shooting} a discretized set of incident rays toward the target and \emph{bouncing} them via specular reflections to capture multiple interactions. This is followed by a PO accumulation step. SBR was introduced in the electromagnetics literature to enable practical radar cross section (RCS) prediction for geometries that were difficult to treat with full-wave tools, such as open-ended cavities~\cite{ling_shooting_1986}, and has since been developed and applied broadly as a ray-based alternative in the HF regime~\cite{sefi_cem_2005}. While foundational concepts date back to classical works, recent advancements in GPU architectures and parallel hierarchies have drastically revitalized ray-based methods for contemporary CEM applications.

SBR requires repeated ray–surface intersection queries across large ray ensembles and multiple reflections. In addition, the number of ray–triangle tests grows rapidly with ray count and bounce depth. As a result, naïve all-against-all ray–triangle testing becomes infeasible at the sampling densities needed for stable PO accumulation. This work targets that dominant bottleneck by accelerating ray–triangle intersection with a bounding volume hierarchy (BVH), a hierarchical spatial data structure that reduces intersection cost by culling large portions of the mesh using bounding boxes~\cite{meister_survey_2021}.

When edge and shadowing effects are important, SBR is often augmented with high-frequency diffraction theories such as GTD, and refinements such as UTD or PTD that improve behavior near shadow boundaries and enforce continuity between illuminated and shadowed regions \cite{keller1962geometrical,kouyoumjian_asymptotic_1965,ahluwalia_uniform_1968,kouyoumjian_uniform_1974,ufimtsev2007fundamentals,breinbjerg_higher_1992}. From a numerical perspective, the PO surface integral is evaluated through a discretization induced by the incident ray grid, so insufficient sampling can introduce artifacts in the accumulated field. However, in high-fidelity multiple-reflection simulations on complex triangulated targets, the dominant practical challenge is typically computational.

The goal of this work is to develop and assess a computationally efficient SBR method for HF electromagnetic backscattering in the presence of complex geometry and multiple reflections. Specifically, we \emph{(i)} formulate a ray-tube discretization of the PO surface integral driven by a controlled incident-ray sampling rule, \emph{(ii)} employ a BVH to reduce the cost of repeated ray–triangle queries during multi-bounce transport, and \emph{(iii)} organize the procedure as a trace–integrate pipeline in which ray traversal produces compact intersection data that is subsequently consumed by PO field evaluation. The method is accelerated on NVIDIA and AMD GPUs and parallelized with MPI by distributing angular sweep parameters across ranks, enabling large parameter studies for electrically large targets. We validate accuracy against analytical Mie solutions for a perfectly conducting sphere and demonstrate applicability to a complex aircraft configuration for monostatic radar cross-section prediction.

\section{Related work}
\label{sec:related}
\noindent\textbf{Ray-based high-frequency electromagnetics.} \hspace{1mm} Computational treatments of PO and related integral evaluations are widely documented in the CEM literature~\cite{mittra_computational_2014}. Shooting and bouncing rays (SBR) is a representative GO+PO technique that combines multi-reflection ray transport with a PO accumulation step to predict quantities such as radar cross section. The method was introduced in the electromagnetics literature to enable RCS analysis of geometries that were challenging for full-wave methods at the time, including open-ended cavities~\cite{ling_shooting_1986}, and later developments positioned ray-based approaches as practical tools for HF scattering and related applications~\cite{sefi_cem_2005}.

Beyond RCS, ray-based solvers and hybridizations appear in problems such as electromagnetic coupling and bidirectional coupling analysis~\cite{taygur_computation_2018,taygur_bidirectional_2018}, propagation and channel modeling~\cite{yun_ray_2015,schiller_gpu_2015,kasdorf_parallel_2024}, antenna placement and design-space exploration~\cite{gomez_accelerated_2023}, and mixed full-wave/ray-tracing strategies that use ray methods where appropriate while retaining full-wave fidelity in subregions~\cite{purcell_combination_2002}.

A known modeling limitation of classical GO+PO SBR is the absence of diffraction mechanisms, which can be important near edges, tips, or shadow boundaries. When needed, diffraction effects are incorporated via HF corrections such as the geometrical theory of diffraction (GTD)~\cite{keller1962geometrical} and refinements such as the uniform theory of diffraction (UTD)~\cite{kouyoumjian_asymptotic_1965,ahluwalia_uniform_1968,kouyoumjian_uniform_1974} or the physical theory of diffraction (PTD)~\cite{ufimtsev2007fundamentals,breinbjerg_higher_1992}. These extensions introduce additional transport rules (e.g., edge-launched rays) while preserving the overall computational pattern of transport followed by accumulation.
\\
In broader CEM workflows, ray-based methods also appear as sub-components in coupling analysis \cite{taygur_computation_2018,taygur_bidirectional_2018} and in design and optimization loops where many configurations must be evaluated \cite{gomez_accelerated_2023}.

\vspace{1mm}
\noindent \textbf{Parallel and heterogeneous acceleration}. \hspace{1mm} To handle the massive ray batches required for complex scenes, recent work has pivoted toward GPU-enabled acceleration. Particularly, hardware-accelerated ray tracing has been leveraged to achieve high-throughput in propagation modeling~\cite{schiller_gpu_2015} and to enable the rapid iteration required for large-scale optimization loops~\cite{gomez_accelerated_2023}. These implementations are often critical when ray-based methods serve as sub-components in hybrid full-wave strategies~\cite{purcell_combination_2002}.

\section{Problem Formulation and Shooting and Bouncing Rays Model}
\label{sec:formulation}
Backscattering analysis for electrically large PEC objects ($D \gg \lambda$) necessitates asymptotic methods, as full-wave discretizations become computationally prohibitive. This work focuses on HF scattering with the goal of computing far-field backscatter, such as monostatic RCS, over large angular parameter sweeps. To bypass the memory constraints of global field solvers, we employ a Shooting and Bouncing Rays (SBR) framework. This replaces volumetric discretization with a transport-and-accumulation procedure: tracing large ray ensembles through the geometry, recording interactions, and coherently integrating surface contributions to compute the total scattered response.

 Time-harmonic fields are conventionally used to represent far-field solutions, which admit an eikonal (Wentzel--Kramers--Brillouin) formulation~\cite{trager_springer_2007,peatross_physics_2023,balanis_balanis_2023}.
\begin{equation}
    \mathbf{E}(\mathbf{r}) = \mathbf{e}(\mathbf{r}) e^{-j k_0 L(\mathbf{r})},
    \qquad
    \mathbf{H}(\mathbf{r}) = \mathbf{h}(\mathbf{r}) e^{-j k_0 L(\mathbf{r})},
    \label{eq:em1}
\end{equation}
where $\mathbf{e}(\mathbf{r})$ and $\mathbf{h}(\mathbf{r})$ are slowly varying complex amplitudes, $L(\mathbf{r})$ is the optical path length (eikonal), and $k_0 = 2\pi/\lambda_0$ is the free space wavenumber with $\lambda_0$ the free space wavelength.

The phase is $\Phi(\mathbf{r}) = k_0 L(\mathbf{r})$. Substituting this ansatz into Maxwell's equations, for $\lambda\to0$, and collecting leading order terms yields the eikonal equation
\begin{equation}
    |\nabla L(\mathbf{r})|^2 = n^2(\mathbf{r}),
    \qquad
    n(\mathbf{r}) = \sqrt{\mu_r(\mathbf{r})\,\varepsilon_r(\mathbf{r})},
    \label{eq:em2}
\end{equation}
where $n(\mathbf{r})$ is the refractive index, $\mu(\mathbf{r})=\mu_r(\mathbf{r})\mu_0$ is the permeability, $\varepsilon(\mathbf{r})=\varepsilon_r(\mathbf{r})\varepsilon_0$ is the permittivity.

Rays are curves $\mathbf{r}(s)$ normal to the wavefronts (level sets of $L$) and satisfy the ray equation
\begin{equation}
    \frac{d}{ds}\!\left( n \frac{d \mathbf{r}}{ds} \right) = \nabla n,
    \label{eq:em3}
\end{equation}
with $s$ a path parameter (often arc length in homogeneous media). In homogeneous media ($\nabla n = 0$), rays are straight lines. In gradient index media they curve, and a common numerical strategy is to approximate curved trajectories with piecewise linear segments. An equivalent Hamiltonian form that is often used for numerical integration is
\begin{equation}
    \frac{d \mathbf{r}}{dt} = \mathbf{p},
    \qquad
    \frac{d \mathbf{p}}{dt} = \frac{1}{2} \nabla n^2(\mathbf{r}).
    \label{eq:em4}
\end{equation}

Geometrical optics provides ray trajectories and local incidence information, but the scattered field is obtained by accumulating contributions over the illuminated surface. In physical optics, the induced surface current on a PEC surface is approximated by the tangent plane approximation
\begin{equation}
    \mathbf{J}_s \approx
    \begin{cases}
        2\, \hat{n} \times \mathbf{H}_{\mathrm{inc}}, & \text{illuminated region}, \\
        0, & \text{shadowed region},
        \label{eq:em5}
    \end{cases}
\end{equation}
where $\hat{n}$ is the unit outward surface normal and $\mathbf{H}_{\mathrm{inc}}$ is the incident magnetic field at the surface.

In the far-field, the scattered electric field follows from the Stratton--Chu formulation~\cite{stratton_em_theory_1941}
\begin{equation}
    \mathbf{E}_s(\mathbf{r}) \approx
    \frac{-j\omega\mu}{4\pi r} e^{-jkr}
    \int_{S}
    \left[
        \mathbf{J}_s(\mathbf{r}') - \big(\mathbf{J}_s(\mathbf{r}') \cdot \hat{r}\big)\hat{r}
    \right]
    e^{jk\mathbf{r}' \cdot \hat{r}} \, dS',
    \label{eq:em6}
\end{equation}
where $\mathbf{r}$ is the observation point, $r=|\mathbf{r}|$, $\hat{r}=\mathbf{r}/r$, $\mathbf{r}'$ is a point on the surface $S$, and $k=\omega\sqrt{\mu\varepsilon}$ is the wavenumber in the surrounding medium.

Eqs.~\ref{eq:em1}--\ref{eq:em6} summarize standard high-frequency asymptotic CEM derivations. The assumption $\lambda \to 0$ in the eikonal derivation implies that the characteristic dimension $D$ is many orders of magnitude larger than the wavelength, allowing wave transport to be approximated by localized ray trajectories without requiring physical wavelengths of exactly zero. Our primary methodological novelty lies in mapping the continuous Stratton--Chu integral (Eq.~\ref{eq:em6}) into a computationally efficient, parallel discrete SBR summation (Eq.~\ref{eq:po_sum}) accelerated by bounding volume hierarchies, valid for PEC surfaces and monostatic backscatter\footnote{A bistatic scattering calculation would require a visibility function to account for shadow regions between the reflection points and the final receiver. In a monostatic configuration this can be omitted, since the predominantly scattering surfaces are by assumption visible to the co-located launcher and receiver.}.
\begin{equation}
    A =
    \sum_{i=1}^{N}
    \frac{j k \Delta A}{4\pi}\,
    2\,(\hat{n}_i \cdot -\hat{k}_{\mathrm{inc}})\,
    \Gamma^{N_i}\,
    e^{-j 2 k R_i},
    \label{eq:po_sum}
\end{equation}
where $\hat{n}_i$ is the surface normal at the first interaction, $\hat{k}_{\mathrm{inc}}$ is the incident propagation direction, $R_i$ is the accumulated optical path length along the ray including reflections, $N_i$ is the number of reflections, and $\Gamma$ is a reflection coefficient that may depend on polarization and incidence angle.

In the SBR method, a discretized set of rays is launched toward the target; each ray undergoes specular reflections while the solver records the quantities needed for field evaluation. The Stratton--Chu surface integral is then evaluated by accumulating the physical optics contribution from each ray's footprint. We discretize that integral by ray tubes, treating each ray as representing an area element $\Delta A$ on an equivalent incident aperture. Rather than tracking ray tube divergence through spreading factors, we control integration error through the density of the incident ray grid, converting accuracy requirements directly into a resolution requirement on ray spacing. It is important to note that the $D \gg \lambda$ assumption applies to the macroscopic object geometry, not to individual mesh triangles. Since rays act as infinitesimally thin test probes, the formulation remains valid even when the local triangle area differs from $\Delta A$, provided the global surface is adequately sampled.

This formulation makes the dominant computational cost explicit: for each ray and each reflection bounce, one must find the next intersection on a complex triangulated surface and accumulate a contribution to the field sum. In high-fidelity multiple-reflection settings this repeated ray--triangle intersection workload dominates, motivating the acceleration structures and pipeline developed in the following sections.

When diffraction effects are important, the formulation can be augmented with theories such as GTD, and with refinements such as UTD and PTD that improve behavior near shadow boundaries and enforce continuity between illuminated and shadowed regions~\cite{carluccio_efficient_2008}. These extensions introduce additional ray generation mechanisms and more complex transport rules, but they fit naturally within the same shooting-and-bouncing framework.

\begin{figure}[t]
    \centering
    \def\imgheight{5cm} 
    \begin{minipage}[t]{0.48\textwidth}
        \centering
        \includegraphics[height=\imgheight, width=\linewidth, keepaspectratio, clip=true, trim = 0 -0.5cm 0 0]{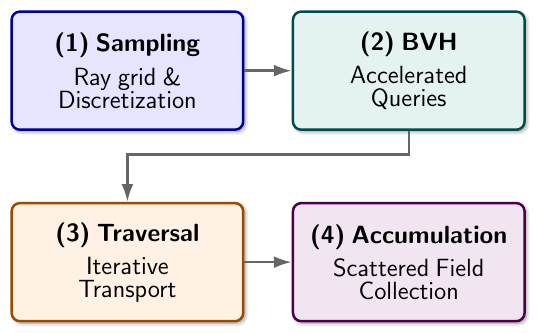}

        \vspace{-1mm}
        \caption{Summary of the SBR pipeline. Rays are launched from the orthographic grid, the BVH accelerates hit detection and traversal by reducing the search space. The scattered field contributions are then accumulated to compute the RCS of the object.}
        \label{fig:pipeline}
    \end{minipage}
    \hfill
    \begin{minipage}[t]{0.48\textwidth}
        \centering
        \includegraphics[height=\imgheight, width=\linewidth, keepaspectratio, clip=true, trim=0.5cm 0.9cm 1cm 0.5cm]{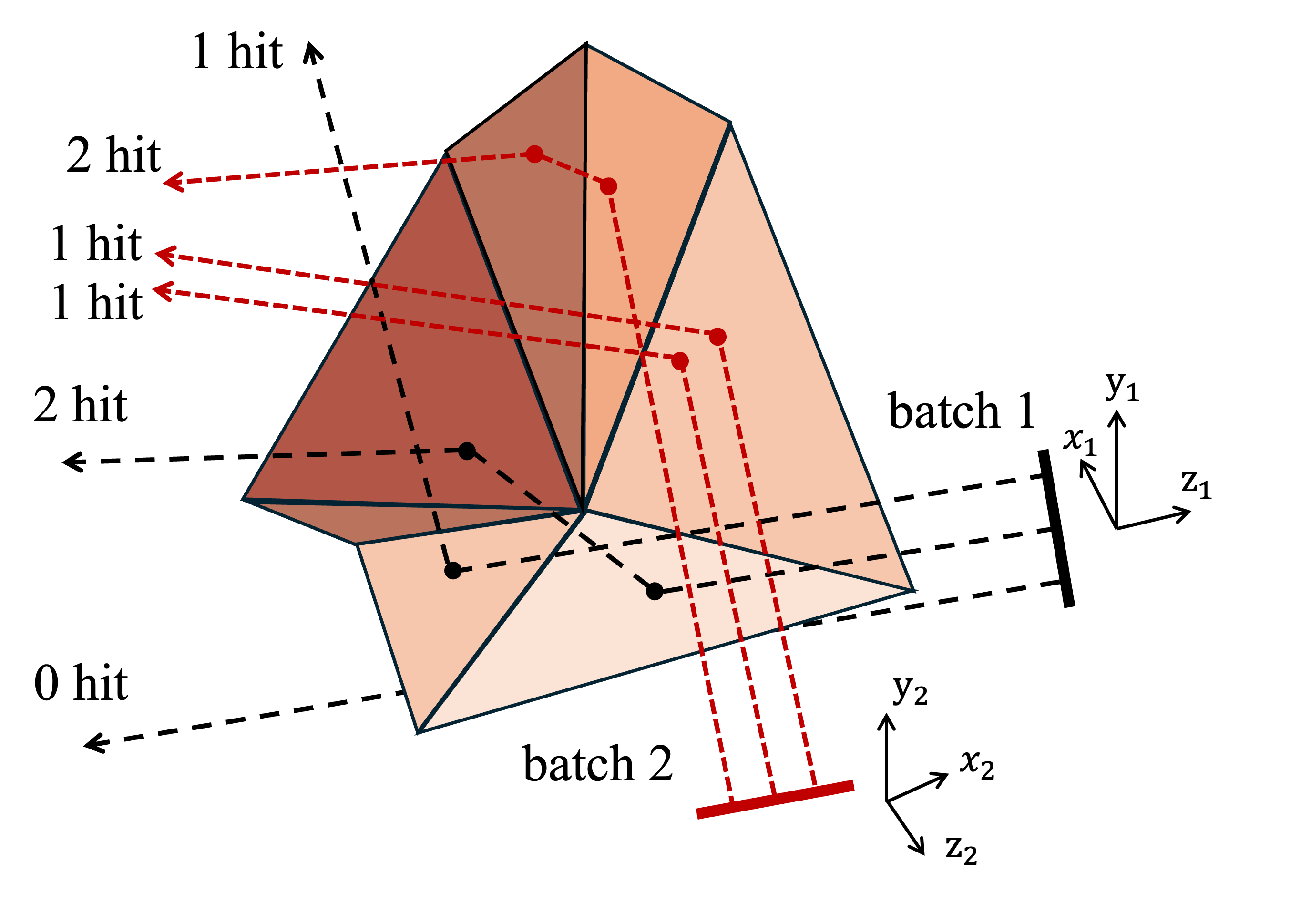}

        \vspace{-1mm}
        \caption{Each ray is transported independently. At the final hit, the distance traveled, the outgoing direction, the surface normal, and the total hit count are recorded. 'Batch 1' and 'Batch 2' illustrate separate ray bundles launched during the angular sweep.}
        \label{fig:ray_hits}
    \end{minipage}
\end{figure}

\section{Methodology}
\label{sec:method}
\noindent This section describes the algorithmic structure used to evaluate monostatic high-frequency scattering with shooting and bouncing rays (SBR) as summarized in Fig.~\ref{fig:pipeline}.

In the first phase, for each incident direction $(\theta,\phi)$ we approximate the incident field by an orthographic bundle of rays with propagation direction $\hat{k}_{\mathrm{inc}}$. The rays originate from a virtual aperture plane orthogonal to $\hat{k}_{\mathrm{inc}}$ and form a Cartesian grid with spacing $\Delta s$. Each ray represents a ray tube with area weight $\Delta A = \Delta s^2$.

If $B$ denote an axis aligned bounding box of the target mesh in world coordinates, for a given incident direction $\hat{k}_{\mathrm{inc}}$, we define an orthonormal basis $\{\hat{u},\hat{v},\hat{k}_{\mathrm{inc}}\}$ where $\hat{u}$ and $\hat{v}$ span the aperture plane. The vectors $\hat{u}$ and $\hat{v}$ correspond to the $x$ and $y$ directions in Fig.~\ref{fig:ray_hits}, while the $\hat{k}_{\mathrm{inc}}$ corresponds to the opposite of the $z$-axis to follow standard ray tracing conventions. Projecting the eight corners of $B$ onto $\hat{u}$ and $\hat{v}$ yields extents
\begin{equation}
    L_u = \max_{c \in \mathcal{C}(B)} (c \cdot \hat{u}) - \min_{c \in \mathcal{C}(B)} (c \cdot \hat{u}), \qquad
    L_v = \max_{c \in \mathcal{C}(B)} (c \cdot \hat{v}) - \min_{c \in \mathcal{C}(B)} (c \cdot \hat{v})  ,
    \label{eq:L1}
\end{equation}
with $\mathcal{C}$ being the set of corners. We choose the aperture size to cover these projected extents with a small margin factor $(1+\eta)$:
\begin{equation}
    \tilde{L}_u = (1+\eta) L_u, \qquad \tilde{L}_v = (1+\eta) L_v.\label{eq:L2}
\end{equation}
The number of rays is then
\begin{equation}
    N_u = \left\lceil \frac{\tilde{L}_u}{\Delta s} \right\rceil, \qquad
    N_v = \left\lceil \frac{\tilde{L}_v}{\Delta s} \right\rceil, \qquad
    N_{\mathrm{rays}} = N_u N_v.
    \label{eq:L3}
\end{equation}
Eqs.~\ref{eq:L1}--\ref{eq:L3} therefore define the projected aperture dimensions and corresponding ray count required for a given incident direction. 

The PO integral is evaluated through a discrete sum induced by the ray grid. If $\Delta s$ is too large relative to the wavelength, the discrete sum undersamples the phase variation across the surface and produces nonphysical oscillations and incorrect null structure in the RCS pattern. In our experiments this under resolution manifests as spatial aliasing in the discretized PO accumulation rather than floating point precision limitations. To control this failure mode we enforce a conservative sampling rule $\Delta s \le {\lambda} / {5}$. This converts accuracy requirements into a resolution requirement on the incident ray grid and provides a simple knob for numerical stability across frequency sweeps. This criterion implies that at least five samples per wavelength are required to accurately capture the rapidly oscillating phase terms. As frequency increases, this dictates a quadratic $O(\Delta s^{-2})$ growth in ray count, establishing a direct tradeoff between physical integration fidelity and computational cost.

\begin{figure}[t]
\centering
\includegraphics[width=0.99\linewidth, 
clip = true, trim = 0 0cm 0 0
]{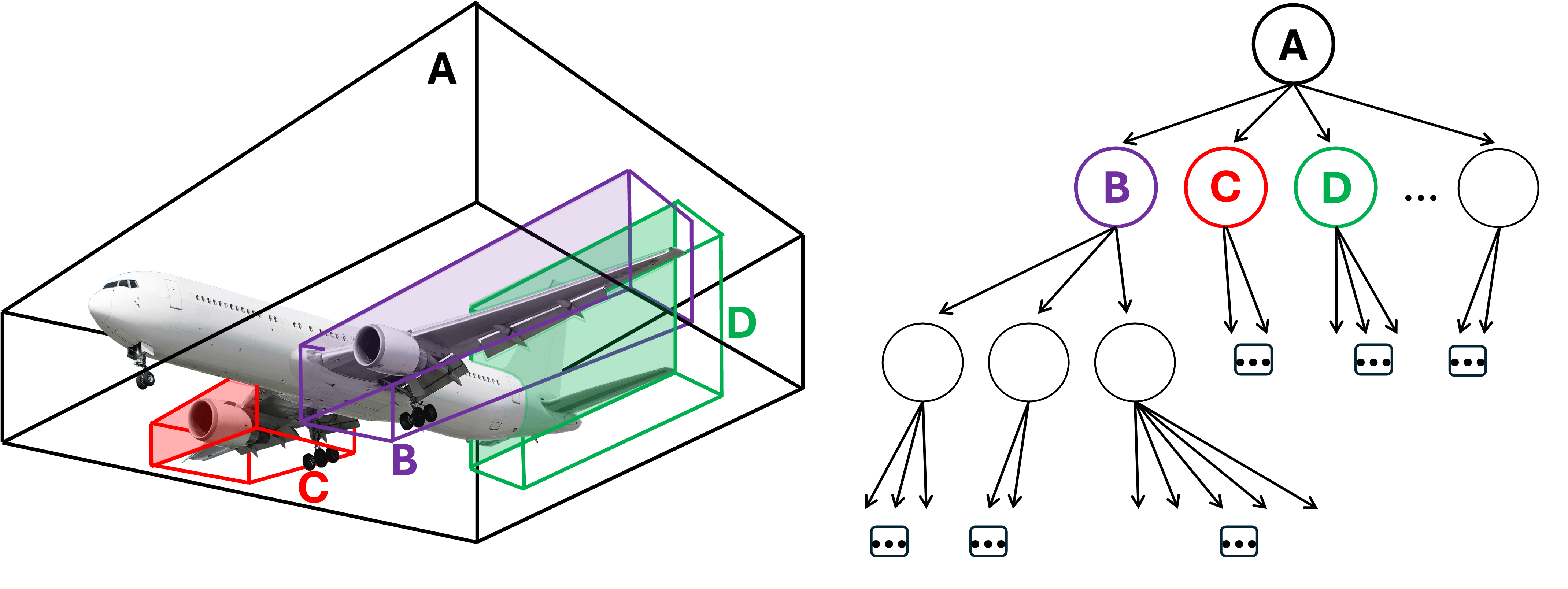}
\caption{Visualizing a BVH constructed using the Binned SAH method. Each bounding AABB is created and recursively subdivided until the finest geometric details are represented. This hierarchical structure accelerates hit detection and traversal by enabling efficient traversal through the BVH tree. The letters A, B, C, and D represent hierarchical spatial partitions of the geometry, where 'A' is the global bounding box, and subsequent letters denote localized sub-volumes (e.g., wings, tail).}
\label{fig:bvh_illustrated}
\end{figure}

The dominant cost in multi reflection SBR is repeated ray triangle intersection. We reduce this massive intersection search space with a BVH built over the mesh triangles (as visually summarized in Fig.~\ref{fig:bvh_illustrated}). Each BVH node stores an axis aligned bounding box (AABB) and either references two child nodes (internal node) or a contiguous range of triangles (leaf node). BVH construction is performed once per mesh and reused across all angles. We support two complementary splitting rules: (i) \emph{Median split.} Triangles are split by the median of centroid coordinates along the longest axis of the current node AABB. This produces balanced trees quickly. (ii) \emph{Binned SAH.} For complex geometry we use a binned surface area heuristic (SAH) that approximates the expected traversal cost. For a candidate split of a parent node $P$ into children $L$ and $R$, the SAH objective is
\begin{equation}
    C_{\mathrm{SAH}}(P \rightarrow L,R) = C_T + \frac{SA(L)}{SA(P)} N_L C_I + \frac{SA(R)}{SA(P)} N_R C_I,
\end{equation}
where $SA(\cdot)$ denotes AABB surface area, $N_L$ and $N_R$ are triangle counts, and $C_T$ and $C_I$ are traversal and intersection costs (treated as constants). In practice, we evaluate this objective over a small number of bins per axis and choose the minimum cost split.

BVH build is a recursive partitioning process (detailed in Algorithm~\ref{alg:bvh_build}) that naturally admits task parallelism. We subdivide nodes until a leaf threshold $N_{\mathrm{leaf}}$ (or depth limit) is reached. Subproblems are independent once the triangle ranges are determined, so node builds can be scheduled as tasks. A global node pool can be allocated with an atomic counter to avoid locking during concurrent node creation. A preorder layout for nodes improves locality during traversal because parent and near child nodes are visited together. Triangles are stored in a compact structure of arrays or tightly packed array of structs; the key design requirement is that leaves reference contiguous triangle ranges.

\begin{figure}[t]
\centering
\begin{minipage}[t]{0.4\linewidth}
\begin{algorithm}[H]
\caption{\\ BVH construction (selectable split)}
\label{alg:bvh_build}
\begin{algorithmic}[1]
\small
\makeatletter\setlength{\ALG@tlm}{-0.35em}\makeatother
\vspace{0.53mm}
\Require Triangles $T$, threshold $N_{\mathrm{leaf}}$, split rule (\textsc{Median}/\textsc{SAH})
\Ensure Linearized arrays \texttt{nodes}, \texttt{tris}
\State Compute centroids and global AABB; initialize root with $T$
\State \textsc{Build}$(\mathrm{root})$
\State Linearize nodes (preorder); store triangle ranges
\Function{Build}{$n$}
    \If{$|T(n)| \le N_{\mathrm{leaf}}$}
        \State Mark $n$ as leaf; \Return
    \EndIf
    \State Choose split axis/position; partition $T \to T_L, T_R$
    \State Create children $L, R$ with AABBs
    \State Parallel: \textsc{Build}$(L)$, \textsc{Build}$(R)$
\EndFunction
\end{algorithmic}
\end{algorithm}
\end{minipage}
\hspace{7mm}
\begin{minipage}[t]{0.5\linewidth}
\begin{algorithm}[H]
\caption{\\ Multi-reflection SBR traversal}
\label{alg:multibounce}
\begin{algorithmic}[1]
\small
\makeatletter\setlength{\ALG@tlm}{-0.35em}\makeatother
\Require BVH \texttt{nodes}, \texttt{tris}, ray $(\mathbf{o},\hat{d})$, max bounces $B_{\max}$
\Ensure Valid flag, $\hat{n}_0$, path length $R$, bounces $N$
\State $R \gets 0$, $N \gets 0$; \texttt{valid} $\gets$ \textbf{false}
\For{$b = 1$ to $B_{\max}$}
    \State $(\texttt{hit}, t, \hat{n}) \gets$ \textsc{Intersect}$(\mathbf{o},\hat{d},\texttt{nodes},\texttt{tris})$
    \If{$\neg \texttt{hit}$}
        \State \textbf{break}
    \EndIf
    \State $\mathbf{x} \gets \mathbf{o} + t\hat{d}$
    \State $R \gets R + t$, $N \gets N + 1$
    \If{$N = 1$}
        \State $\hat{n}_0 \gets \hat{n}$; \texttt{valid} $\gets$ \textbf{true}
    \EndIf
    \State $\hat{d} \gets \hat{d} - 2(\hat{d}\!\cdot\!\hat{n})\hat{n}$
    \State $\mathbf{o} \gets \mathbf{x} + \epsilon \hat{n}$
\EndFor
\State \Return (\texttt{valid}, $\hat{n}_0$, $R$, $N$)
\end{algorithmic}
\end{algorithm}
\end{minipage}
\end{figure}

Given an incident ray grid and the BVH, the traversal stage {(Algorithm~\ref{alg:multibounce}) propagates each ray through up to $B_{\max}$ specular reflections. Each reflection requires computing the closest ray triangle intersection. We use an iterative BVH traversal with an explicit stack. The traversal performs a standard AABB intersection test to cull nodes and then tests triangles in leaf nodes. The closest hit along the ray is returned. The explicit stack stores node indices. Nodes are pushed in near first order to increase early pruning.

After a hit at point $\mathbf{x}$ with unit normal $\hat{n}$, the ray direction is updated by specular reflection $\hat{d}_{\mathrm{new}} = \hat{d}_{\mathrm{old}} - 2(\hat{d}_{\mathrm{old}} \cdot \hat{n}) \hat{n}$. To avoid immediately re intersecting the same surface due to finite precision arithmetic, we offset the new origin by a small $\epsilon$ along the normal or reflected direction, following common practice in ray based intersection pipelines~\cite{kwon_resolution_2023}. While a BVH does not reduce the intrinsic arithmetic cost of a single ray-triangle intersection test, it drastically reduces the massive intersection search space by culling large portions of the mesh using bounding boxes.

In the \emph{PO integration and trace integrate decomposition} phase, the SBR evaluation decomposes into two stages with different control flow. First, a traversal stage produces a compact hit record per ray: a validity flag, the first hit normal, the total optical path length $R_i$, and the reflection count $N_i$. Importantly, the trace stage is dominated by BVH traversal and triangle tests, and its control flow depends on geometry and reflection depth. Second, the PO accumulation stage evaluates the coherent sum using only valid rays. Here, an ``invalid'' ray is defined as one that entirely misses the target geometry (escaping into free space) or becomes permanently trapped in a cavity without returning to the receiver plane. For a PEC surface and monostatic backscatter we compute the complex amplitude, detailed in Eq.~\ref{eq:po_sum}, and the monostatic RCS is $\sigma = 4\pi |A|^2$, which is evaluated in $\text{m}^2$ and typically expressed in dBsm as $10 \log_{10}(\sigma)$.

Because invalid rays are filtered out, the integrate stage has regular control flow and can be implemented as a reduction over the valid hit records. This separation reduces wasted work compared to computing field contributions inside the traversal loop, and it allows the intersection dominated stage and the arithmetic dominated stage to be optimized independently.

In final stage, a monostatic sweep over $(\theta,\phi)$ is embarrassingly parallel because each incident direction is independent given the fixed geometry. We exploit this by distributing the set of angles across different processes, treating the sweep as a parameter exploration problem. Each process holds a copy of the mesh and BVH and evaluates the trace integrate procedure for its assigned angle subset.

The algorithm exposes simple cost drivers that guide both scaling analysis and practical parameter selection. Binned SAH adds a constant factor due to bin evaluation per node but preserves the same asymptotic behavior in typical settings. The BVH is built once and amortized across all angles. This term typically dominates runtime for electrically large targets because accurate PO evaluation drives $N_{\mathrm{rays}}$ upward and multi reflection effects increase the effective number of intersection queries. The PO accumulation stage is a reduction over valid hits, with a small constant factor for complex arithmetic.

\subsection{Implementation}
\label{subsec:implementation}
\noindent The method is implemented in \texttt{SagittaSBR}~\cite{SagittaSBR}, a modular C++/CUDA code for high-throughput SBR evaluation on heterogeneous systems. It follows a standard offloading model: the host performs geometry I/O, BVH construction, and sweep orchestration, while per-angle work (multi-bounce traversal and PO accumulation) runs on GPUs. The HIP runtime API is then used to also support AMD architectures from a single source base.

At startup, the host loads a triangulated mesh (e.g., \texttt{.obj} or \texttt{.gltf}) and builds a BVH over its triangles using the splitting strategies described earlier. Construction proceeds via recursive partitioning. To reduce wall-clock time for large meshes, subtree builds are issued as OpenMP tasks once triangle ranges are split. Nodes are allocated from a global pool indexed by \texttt{std::atomic} counters to avoid lock contention. The final BVH and triangle data are linearized into contiguous arrays (\texttt{nodes}, \texttt{triangles}) for traversal.

For each MPI rank, these arrays are transferred to GPU memory and reused for all assigned incident angles. During sweeps, per-ray state resides in preallocated device buffers sized to the largest ray grid in the run. The dominant state consists of two 3-vectors (current direction and first-hit normal), the accumulated path length, and the bounce count, i.e., $2\times \texttt{sizeof(vec3)} + \texttt{sizeof(Real)} + \texttt{sizeof(int)}$ per ray.

Geometry data (\texttt{gpuBvhNodes}, \texttt{gpuBvhTris}) is mesh-dependent and shared across rays. Storing nodes and triangles contiguously improves locality and ensures each leaf references a contiguous triangle range. For a fixed incident direction, rays are launched on an orthographic grid and mapped one-to-one to GPU threads. Each thread performs multi-bounce transport by iteratively traversing the BVH to locate the nearest intersection, updating the direction, and accumulating optical path length. Initially, the orthographic launch grid provides excellent ray coherence and L1 cache hit rates. However, after subsequent reflections, rays diverge significantly. To maintain high thread occupancy and avoid register spilling in this divergent regime, the BVH traversal strictly bounds local state using an explicit, fixed-size memory stack. While hardware rasterization is extremely fast for primary visibility, we specifically employ ray tracing to handle the complex, incoherent secondary bounces required for scattering computations.

Traversal uses an explicit stack to avoid recursion and support deep trees robustly. The traversal stage writes a compact hit record (valid flag, first-hit normal, total path length, bounce count), consumed by a second kernel for PO accumulation. Only valid rays contribute to the far-field amplitude. This trace–integrate decomposition separates intersection-dominated transport from arithmetic-dominated reduction and allows the second kernel to operate on a dense, regular buffer of hit records.

\section{Numerical Results}

\noindent To obtain physically accurate results in SBR simulations, the incident ray grid must be carefully configured relative to the target geometry. The virtual aperture must be dimensioned to fully encompass the object's cross-section from all sweep angles by setting its size larger to the largest size of the object. More critically, as explained earlier, the spatial density of the rays must be sufficient to resolve phase variations of the electromagnetic wave. At higher frequencies, numerical artifacts, such as non-physical oscillations or incorrect nulls in the RCS pattern, may appear. While these are occasionally misidentified as precision issues due to the increasingly small phase terms, they are typically the result of spatial aliasing. Specifically, under-resolving the wave physics leads to integration errors in the PO kernel, as the discretized sum fails to accurately represent the continuous surface integral.

\subsection{Validation with PEC sphere}

\begin{figure}[t]
    \centering
    \includegraphics[width=1\linewidth]{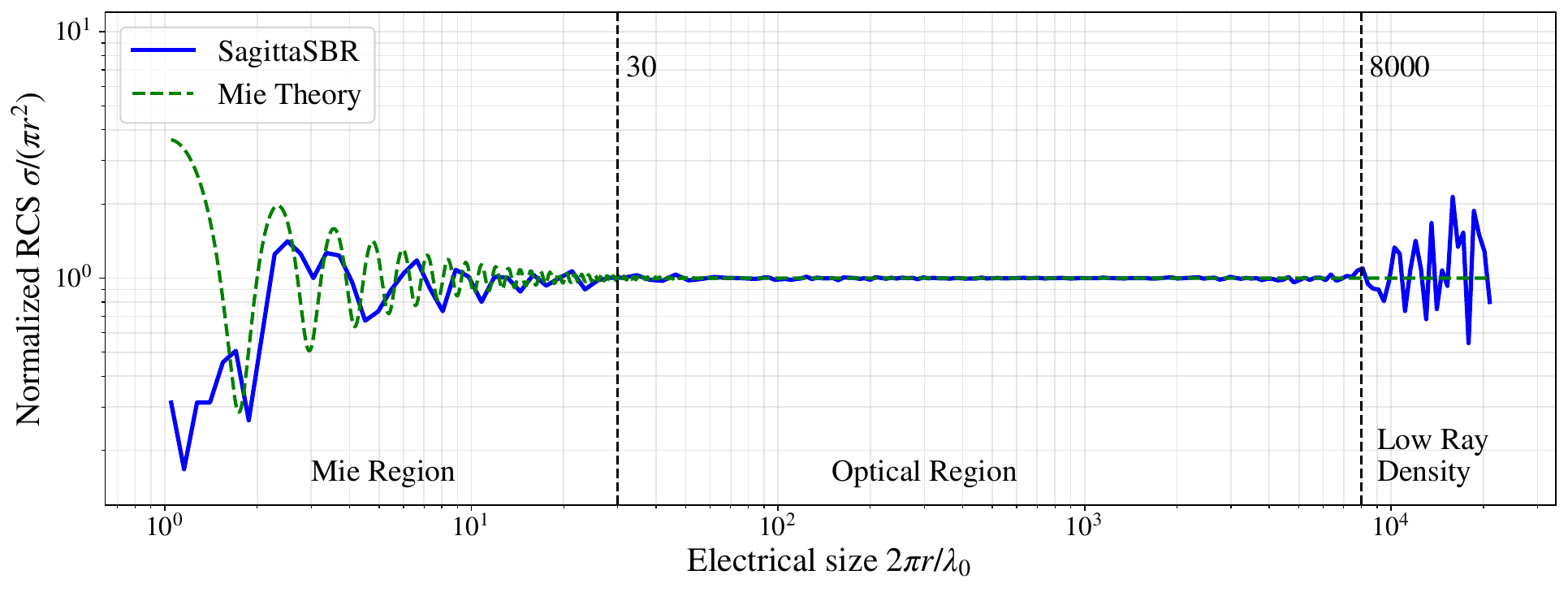}
    \vspace{-2mm}
    \caption{Frequency scan of monostatic RCS for a triangulated PEC sphere, averaged across incident directions. The results highlight the Mie resonance region where ray tracing is inapplicable, followed by the optical region where the method accurately approximates scattering with a standard deviation of $\sim$2.5\%. At higher frequencies, a low ray density region fails to resolve the wavelength, leading to an aliasing instability. The labels $kr=30$ and $kr=8000$ denote the boundary of the optical region and the onset of the aliasing instability, respectively. Here, $\lambda_0$ denotes the free-space wavelength.}
    \label{fig:rcs_sphere}
\end{figure}

\noindent To validate our implementation, we analyzed the computed RCS of a PEC sphere against analytical Mie scattering theory. Electrical size is here defined as $kr = 2\pi r / \lambda$, where $r$ is the sphere radius. At $kr = 1$, Mie resonances dominate and ray tracing methods fail, as wave effects cannot be neglected in this regime. Ray-based approximations become valid as frequency increases, with good agreement typically expected for $kr \gtrsim 30$.

Our simulation results (Fig.~\ref{fig:rcs_sphere}) demonstrate excellent agreement with Mie theory beginning at an electrical size around $kr \approx 30$. This transition marks the boundary between the Mie resonance region, where wave phenomena dominate, and the optical region, where the ray approximation can accurately capture the scattering physics of the triangulated PEC sphere.

In principle, this agreement should extend to arbitrarily high frequency ranges; however, \emph{practical limitations arise from ray undersampling}. When the ray grid becomes insufficiently dense to accurately resolve the wavelength, aliasing introduces numerical artifacts, known as aliasing instabilities. In our simulations, this phenomenon occurs at electrical sizes around $kr \approx 8 \times 10^3$. To explicitly demonstrate the onset of aliasing, the ray grid density was held strictly constant throughout this specific frequency sweep experiment. As the wavelength decreased at higher frequencies, the static $\Delta s$ eventually violated the $\lambda/5$ criterion. These artifacts can be mitigated by increasing ray density at the expense of computational time.

Importantly, the SBR method is not strictly frequency-independent, as the ray sampling must adequately represent the wavelength scale. Nevertheless, the results are highly promising: the SBR method produces accurate predictions across more than two orders of magnitude in electrical size. For a sphere of radius $r = 1$ m, this corresponds to a frequency range from approximately 1 GHz to 300 GHz, with results remaining within $\pm 2\%$ of the analytical solution. These simulations used a configuration file specifying a $22{,}000 \times 22{,}000$ ray grid launched from a square aperture of 2.05~m on each side.

While we anticipated potential precision limitations at high frequencies when operating in single precision (FP32) for the PO kernel, given that the phases become small, the primary constraint on accuracy remains the ray density specified in the configuration file. Ensuring sufficient sampling of the wavelength is essential for maintaining physical fidelity across the frequency sweep. Additionally, no normalization was introduced in the code as wavelength values remain orders of magnitude larger then machine epsilon in FP32. One additional factor to consider is that between different runs of scans, simulation results may vary due to the nondeterministic behavior of floating-point operations. This can lead to reproducibility issues  when running in FP32. To reduce this variance, it is best practice to repeat simulations multiple times and average the results.

\subsection{Radar Cross Section of Large Metallic Objects}

\begin{figure}[t]
    \centering
    \includegraphics[width=0.9\linewidth, clip = true, trim = 0.3cm 0 0.2cm 0]{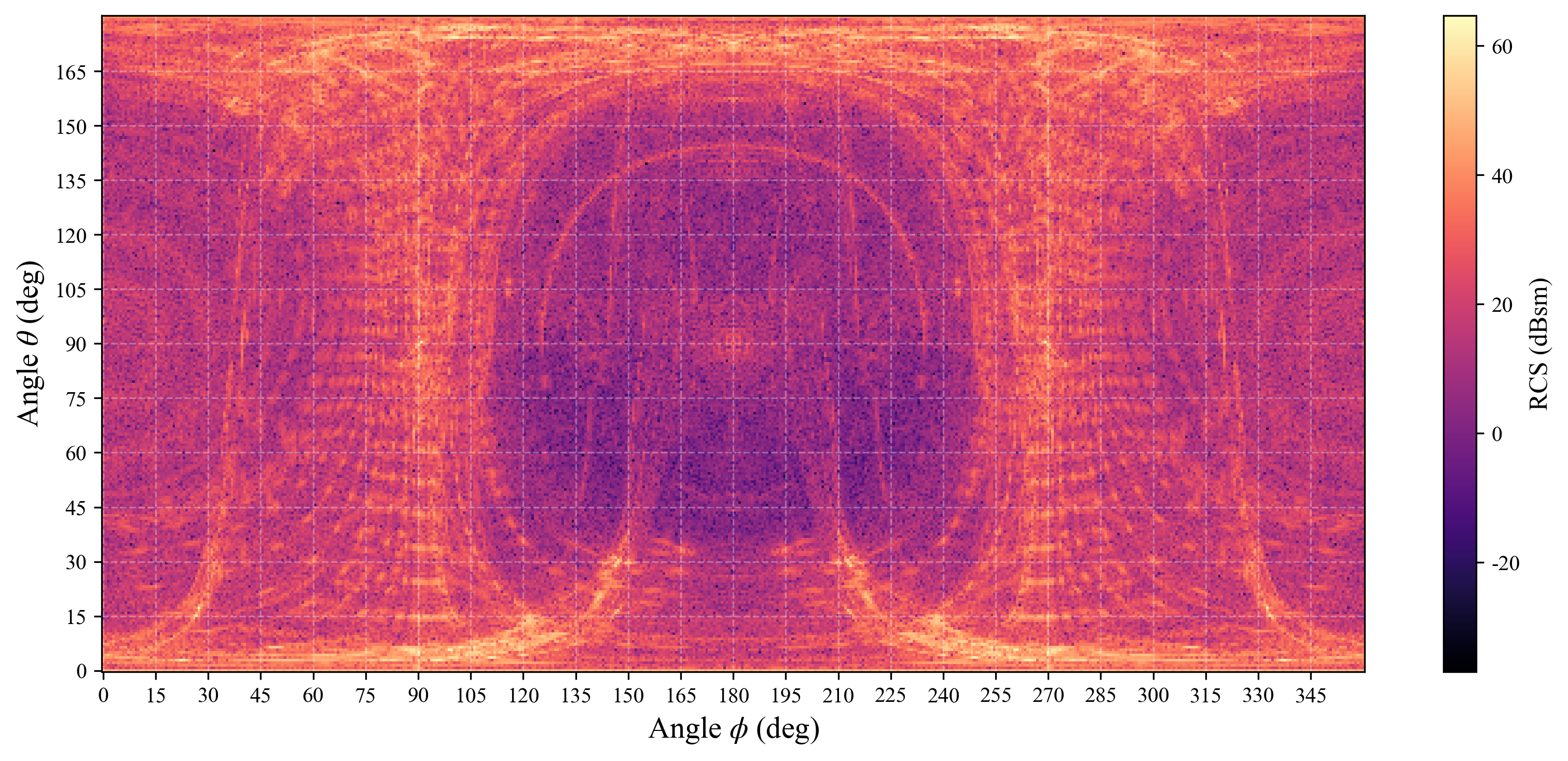}
    \vspace{-2mm}
    \caption{Radar cross-section (RCS) calculation for an A380 aircraft. The simulation uses 500 samples in $\phi$ from $0^\circ$ to $360^\circ$ and 250 samples in $\theta$ from $0^\circ$ to $180^\circ$ at a frequency of 10~GHz ($\lambda \approx 3$~cm). A total of $30{,}000 \times 30{,}000$ rays are launched from a $90 \times 90$~m grid, with an average computation time of 616.12~ms in FP32 per angular position with a total sweep time of $\sim$27 minutes on 8 LUMI nodes. The total ray count of the simulation amounts to $1.125 \times 10^{14}$ rays\protect\footnotemark.}
    \label{fig:rcs_2d}
\end{figure}

\noindent To evaluate the solver performance on electrically large, complex geometries, an A380 aircraft was selected as a primary benchmark. The aircraft is modeled as a PEC object, a standard approximation for metallic structures in microwave regime. The simulation performs a high-resolution, full-spherical monostatic RCS scan. To ensure that the physical optics integral is numerically stable and free of aliasing artifacts, the incident ray grid must be sufficiently dense relative to the wavelength.

For this benchmark at 10~GHz, we utilized a launch grid of $30{,}000 \times 30{,}000$ rays spanning a $90 \times 90$~m aperture. The simulation allowed for a maximum of 100 reflections per ray to capture deep cavity scattering effects, such as those occurring within engine intakes. The results of the full angular sweep are visualized in Figs.~\ref{fig:rcs_2d} and~\ref{fig:a380_and_rcs}. The simulation computed the RCS at 500 azimuthal ($\phi$) and 250 elevation ($\theta$) angles, totaling 125,000 independent observation points. The resulting heatmaps clearly identify high-intensity scattering zones corresponding to broadside specular reflections from the fuselage and wings, as well as complex scattering patterns arising from the tail section and engines.

\begin{figure}[t]
    \centering
    \begin{minipage}[t]{0.37\linewidth}
        \centering
        \includegraphics[
            width=0.999\linewidth,
            clip=true,
            trim=7.5cm 2cm 0.5cm 2cm
        ]{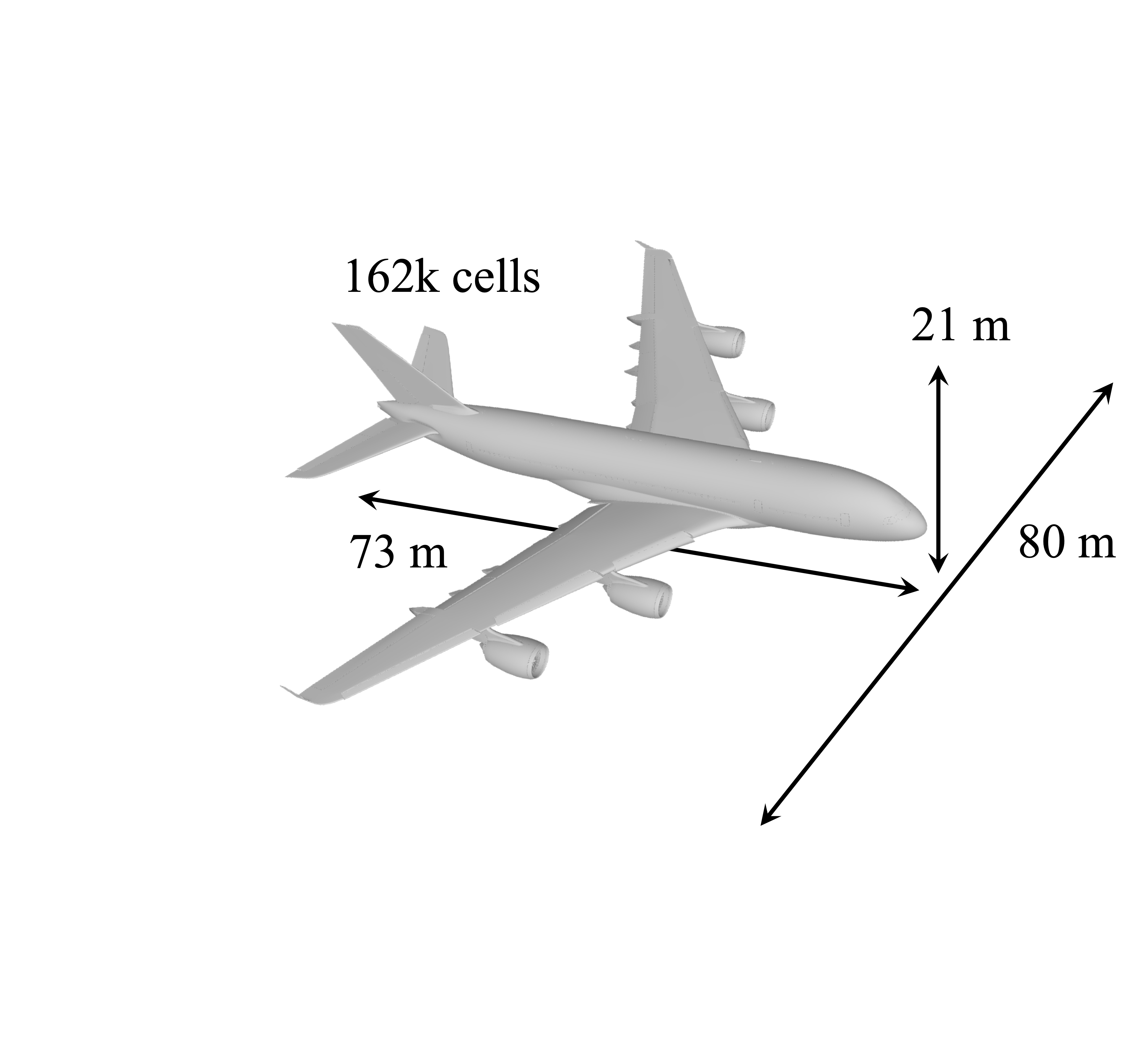}
        
        \vspace{2mm}
        {\small (a)}
    \end{minipage}
    \hspace{7mm}
    \begin{minipage}[t]{0.50\linewidth}
        \centering
        \includegraphics[
            width=0.999\linewidth,
            clip=true,
            trim=0 1cm 0cm 0
        ]{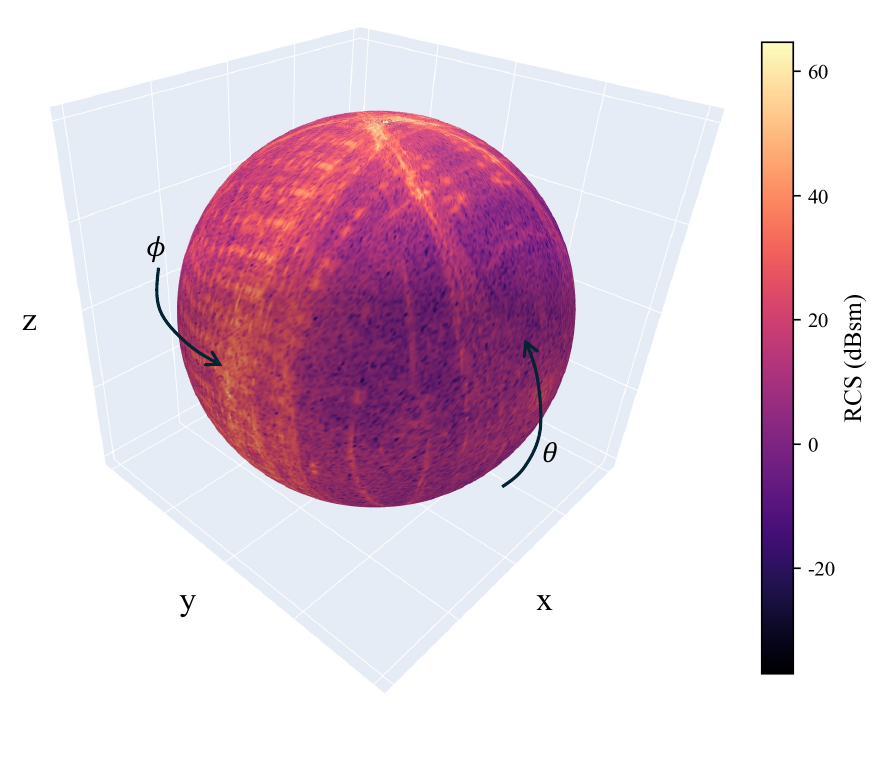} 
        
        \vspace{2mm}
        {\small (b)}
    \end{minipage}
    \vspace{0mm}
    \caption{Aircraft geometry and radar cross section representation. (a) Simulated A380 aircraft model with approximate dimensions of $80 \times 73 \times 21$~m, discretized using 162{,}000 triangular elements,  12.2~MB. (b) Visualization of the RCS revolved around a spherical surface at a large distance from the object.}
    \label{fig:a380_and_rcs}
\end{figure}

\section{Performance Analysis}

\noindent To demonstrate the performance and efficiency of \texttt{SagittaSBR}, we use the LUMI pre-exascale supercomputer, ranked 9th globally in the November 2025 TOP500 list, an HPE Cray EX supercomputer based on AMD EPYC CPUs, AMD Instinct MI250X GPUs, and HPE Slingshot-11 interconnect. The RCS calculation for the A380 aircraft, on a partition of eight LUMI nodes with four AMD MI250X GPUs per node, the average computation time is approximately 616~ms per angular position in FP32. Following this benchmark, the framework was additionally evaluated on a broader range of aircraft geometries to assess consistency and robustness across varying topological features. The computational efficiency of the implementation was evaluated by analyzing both the kernel-level performance on individual devices and the parallel scalability on the LUMI supercomputer.

\footnotetext{For comparison, $1.125 \times 10^{14}$ is roughly the number of grains of sand needed to fill up 4 Olympic pools.} 

To understand the computational cost per angular step, we analyzed the execution trace of the solver. As illustrated in Fig.~\ref{fig:trace}, the simulation pipeline is dominated by the ray-surface intersection tests.

\begin{figure}[t]
    \centering
    \includegraphics[width=0.85\linewidth]{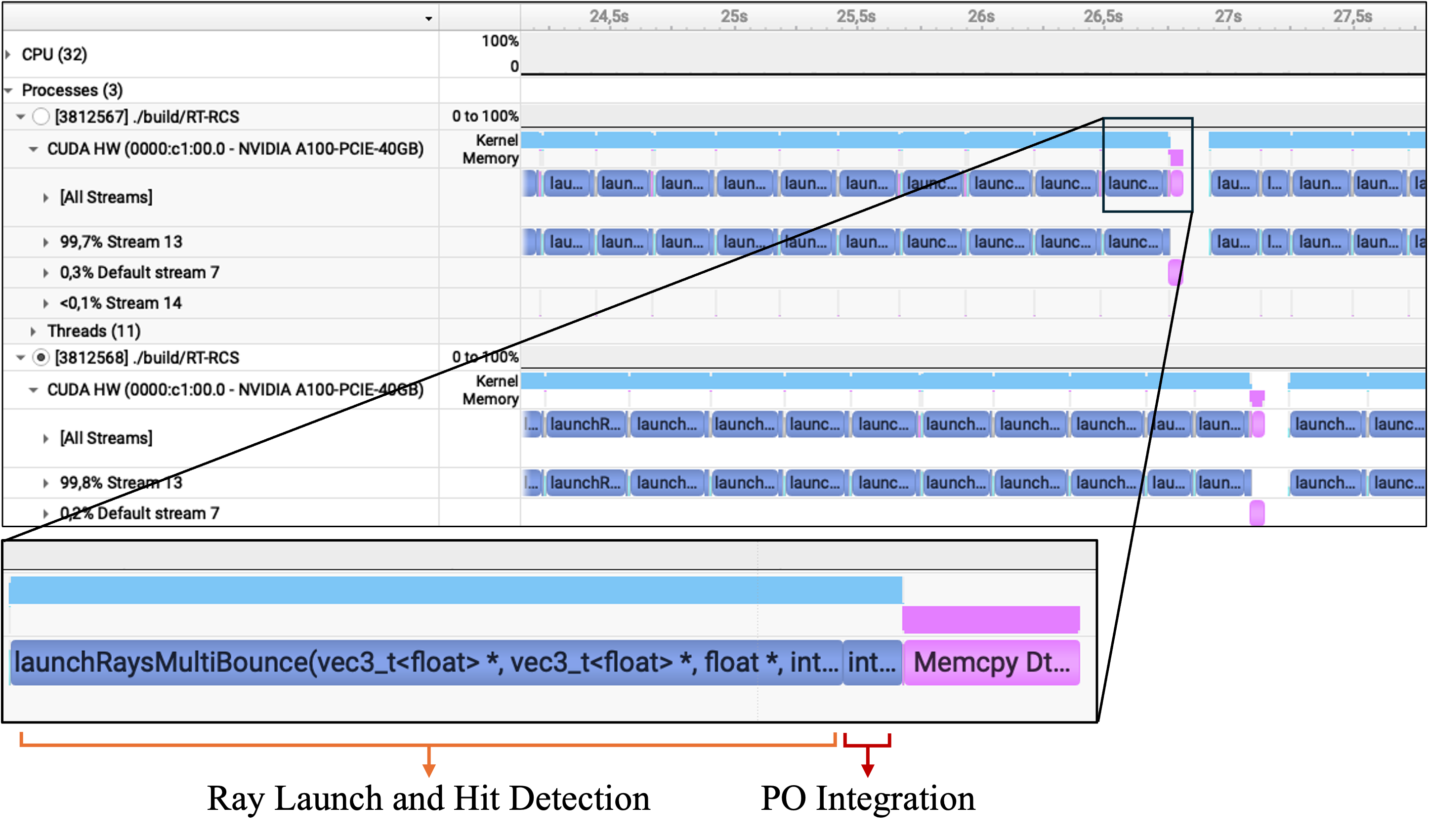}
    \vspace{0mm}
    \caption{Execution trace from NVIDIA Nsight Systems on 2 MPI processes each with one NVIDIA A100. For each scan angle $(\theta_i, \ \varphi_j)$, the simulation is performed by a set of kernels: first, the ray generation kernel that computes intersections, followed by the computation of the Stratton--Chu PO integral and the calculation of the RCS. The intersection kernel requires the highest computational cost.}
    \label{fig:trace}
\end{figure}

\newsavebox{\boxStrong}
\newsavebox{\boxWeak}

\begin{figure}[t]
    \centering
    \def\figHeight{6cm} 

    \sbox{\boxStrong}{%
        \includegraphics[height=\figHeight]{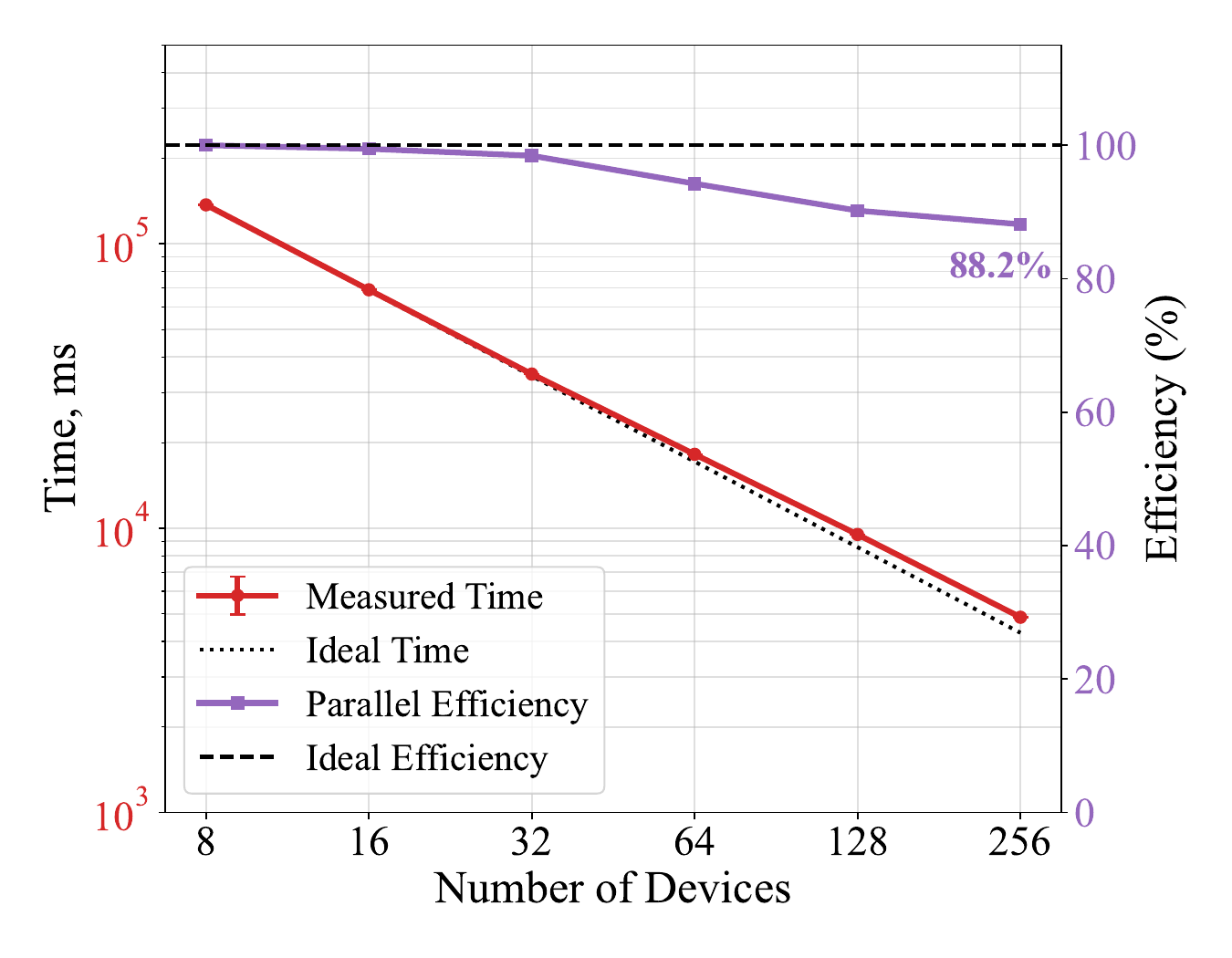}%
    }
    \sbox{\boxWeak}{%
        \includegraphics[height=\figHeight]{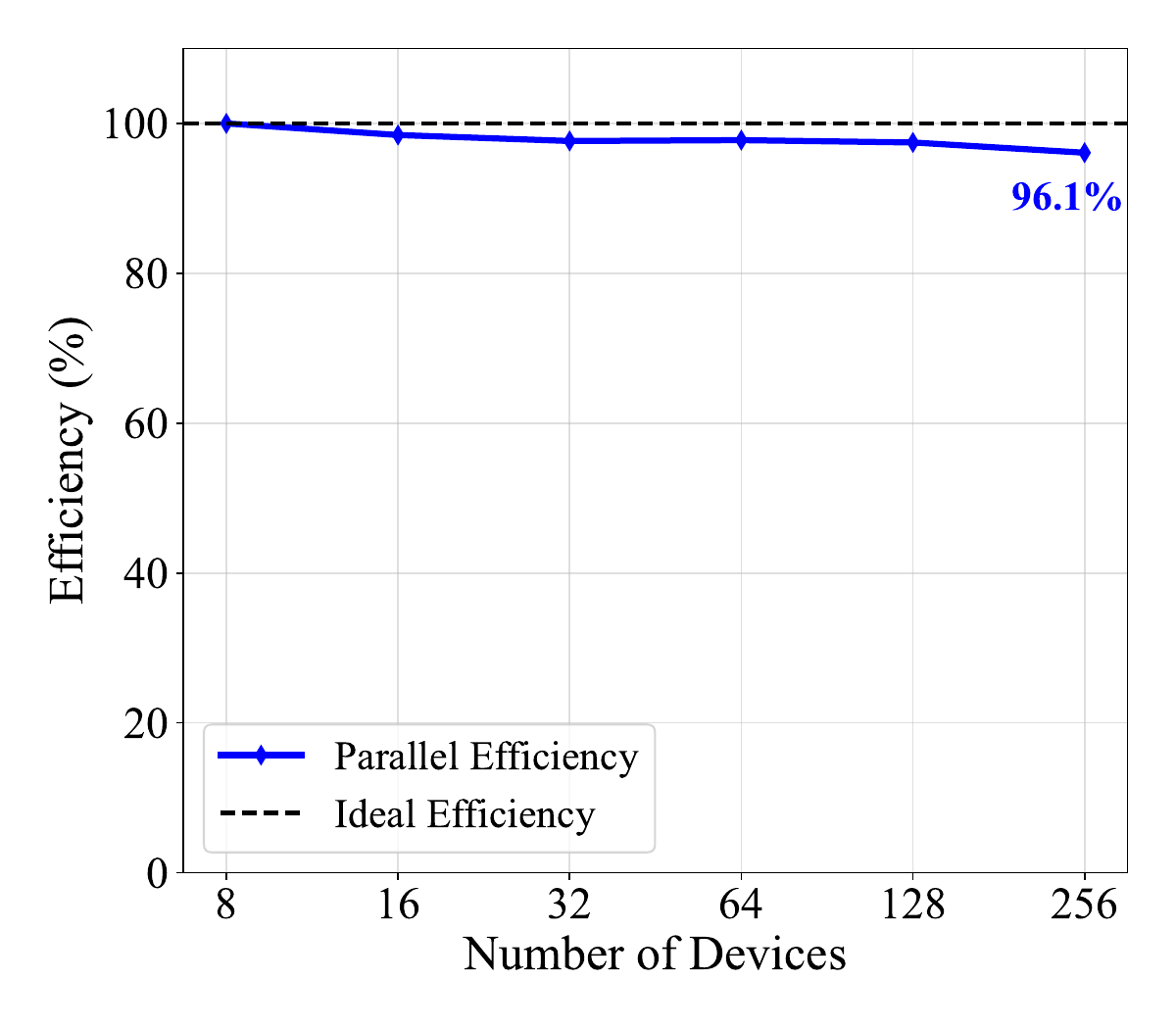}%
    }

    \begin{minipage}[t]{\wd\boxStrong}
        \centering
        \usebox{\boxStrong} 
        \vspace{-3.5mm}
        \\ {\small (a) Strong Scaling}
    \end{minipage}
    \hspace{0.2cm} 
    \begin{minipage}[t]{\wd\boxWeak}
        \centering
        \usebox{\boxWeak} 
        \vspace{-3.5mm}
        \\ {\small (b) Weak Scaling}
    \end{minipage}

    \vspace{0mm}
    \caption{Scaling performance on LUMI, using FP32. On the $x$-axis the total number of GCDs utilized on AMD MI250X GPUs. Each computing node contains 4 physical GPUs, providing a total of 8 GCDs, with 1 MPI process for each GCD. Measurements are averaged over 3 independent trials while launching $45,000 \times 45,000$ rays per scan; very low variance is reported. Strong scaling maintains an efficiency of $\sim88\%$ at the largest scale (256 GCDs), while weak scaling retains $\sim96\%$ efficiency.}
    \label{fig:scalings}
\end{figure}

A significant factor influencing performance is the variation in the number of active rays. Depending on the incident angle relative to the object's profile, the percentage of rays that actually intersect the target can be low, sometimes reaching a range of 5-10\%. While it is possible to implement an optimized launching platform to select rays before launch, this introduces significant tuning overhead. Therefore, a dense, uniform grid is employed to ensure robustness across all angles. Given that often only 5-10\% of launched rays intersect the target, performance could be significantly enhanced by implementing early filtering heuristics. For instance, rasterizing a low-resolution bounding silhouette of the object prior to the ray launch could mask out primary rays destined to miss the target, eliminating unnecessary BVH traversal overhead, albeit at the cost of a preliminary rasterization pass.

We further compared the performance impact of floating-point precision across architectures, including a system with two NVIDIA A100 GPUs, as detailed in Table~\ref{tab:time-breakdown}. While the NVIDIA A100 exhibits the expected performance penalty when switching from single to double precision (FP64), leading to a $\sim53\%$ increase, results indicate that the AMD MI250X shows an inverse behavior, where FP64 execution is approximately $20\%$ faster than FP32. This suggests that the compute units on the MI250X are highly optimized for 64-bit operations for this specific instruction mix.

When selecting precisions, numerical accuracy must be weighed alongside throughput. While FP32 introduces a slight phase accumulation error ($\sim$2--3\% deviation) after 100 reflections due to catastrophic cancellation in the $e^{-j2kR_i}$ term, it is generally acceptable on logarithmic RCS (dBsm) scales. FP64 eliminates this error entirely but incurs the aforementioned penalty on NVIDIA architectures. Consequently, the optimal precision choice is architecture-dependent: FP32 is favored for NVIDIA A100 to maximize throughput, while FP64 is recommended for AMD MI250X to minimize runtime.

The parallel scalability of the framework was assessed on the LUMI supercomputer using the AMD MI250X partition. The parallelization strategy distributes the angular sweep parameters across MPI ranks. It is important to highlight that this approach requires each MPI process to hold the full geometry (\texttt{.obj} model) and the ray buffers in memory. Consequently, the simulation size is bounded by the memory capacity of a single GPU, and different hardware capacities prevent direct scaling of the exact same problem size across heterogeneous clusters.

However, this memory constraint is rarely a limiting factor for practical SBR applications. The method is primarily intended for the GHz regime where objects are electrically large ($kr \lesssim 10^5$). For such cases, the ray density required is manageable within modern GPU memory limits. Conversely, if an object requires extreme ray counts due to microscopic details, the SBR approximation itself typically breaks down, and for extremely large objects, visual detection methods surpass radar ones.

\begin{table}[htpb]
    \centering
    \caption{Kernel-time breakdown for A380 scattering simulations ($18{,}000\times18{,}000$ rays) using one MPI process. Statistics evaluated over 10 simulation repetitions, each simulation scans 600 angular positions. \textit{Note}. Performance is compared per allocatable unit: one full NVIDIA A100, one GCD (half-card) of the AMD MI250X.}

    \vspace{2mm}
    \label{tab:time-breakdown}
    \renewcommand{\arraystretch}{1.1} 
     \setlength{\tabcolsep}{9.pt}
     \vspace{1mm}
    \begin{tabular}{llccc}
        \toprule
        {Device} & {Kernel} & time \texttt{FP32}, ms & time \texttt{FP64}, ms & $\Delta$-time \\
        \midrule
        \multirow{2}{*}{NVIDIA A100 (Full)}
            & Ray Launch   & $98.96  \pm 1.18$ & $151.30 \pm 1.75$ & $+52.9\%$ \\
            & PO Integral  & $6.43   \pm 0.09$ & $6.01   \pm 0.06$ & $-7.0\%$  \\
        \addlinespace[3pt]
        \multirow{2}{*}{AMD MI250X (1 GCD)}
            & Ray Launch   & $249.81 \pm 0.52$ & $198.70 \pm 0.31$ & $-20.5\%$ \\
            & PO Integral  & $4.83   \pm 0.01$ & $7.41   \pm 0.01$ & $+53.4\%$ \\
        \bottomrule
    \end{tabular}

    \vspace{5mm}
\end{table}

As shown in Fig.~\ref{fig:scalings}, the solver demonstrates excellent scalability. Strong scaling efficiency remains at $\sim88\%$ even when utilizing 256 GCDs, while weak scaling efficiency is maintained at $\sim96\%$, indicating that the overhead of MPI communication and BVH management is minimal compared to the ray tracing workload. For the weak scaling tests, a base workload of $45{,}000 \times 45{,}000$ rays per node was maintained. Although the current MPI strategy perfectly distributes angular positions, future implementations must consider workload imbalances across nodes (tail effects) caused by certain incident angles triggering deeply recursive reflections in cavities while others simply reflect off broadside surfaces.


\section{Discussion and Conclusion}
\label{sec:conclusion}
\noindent In this paper, we presented a SBR computational method for rapid monostatic RCS evaluation of electrically large PEC objects, together with an efficient implementation in \texttt{SagittaSBR}. The approach couples multi-reflection geometrical-optics transport with a ray-tube discretization of the PO surface integral. The dominant computational cost in multi-bounce SBR is repeated ray-triangle intersection. We addressed this bottleneck by building a BVH for every MPI process and reusing it across large angular sweeps. Validation against analytical Mie solutions for a PEC sphere confirms that the method reproduces the expected optical-regime behavior.

Large-scale experiments on the LUMI supercomputer demonstrate high parallel efficiency for parameter sweeps on AMD MI250X GPUs. The performance results show that throughput is sensitive to architectural characteristics and precision choices. In particular, we observe distinct single versus FP64 behavior across NVIDIA and AMD devices for the intersection-dominated traversal stage. This shows the importance of architecture-aware tuning in future HF CEM solvers that target heterogeneous exascale systems.

Several extensions can be pursued in the future. On the modeling side, classical GO+PO SBR is limited to specular interactions. Incorporating diffraction corrections (e.g., GTD/UTD/PTD) would improve fidelity in edge- and shadow-dominated configurations. More general interaction models, including dielectric materials and fully polarimetric (vector) scattering, are also of interest. These extensions are algorithmically significant because they introduce additional interaction pathways (e.g., ray spawning) and increase variability in per-ray work, motivating dynamic scheduling strategies and careful memory management. Additionally, since the scattering search space is massive, constructing directionally optimized BVH subtrees that prioritize nodes facing the incident wave could further accelerate intersection queries. On the numerical side, we plan to strengthen robustness by moving toward energy-consistent ray-tube transport.

\vspace{5mm}
\noindent\textbf{Acknowledgments.} This work has received funding from the Swedish Research Council's Research Environment grant (SEE-6GIA 2024-06482).

\FloatBarrier


\printbibliography

\end{document}